\documentclass[10.75pt]{article}
\usepackage{geometry}
\usepackage{makeidx}
\usepackage[T1]{fontenc}
\usepackage[dvipsnames,svgnames,table]{xcolor}
\usepackage{epstopdf}
\usepackage{epsfig}
\usepackage{textcomp}
\usepackage{hyperref}
\usepackage{amsmath}
\usepackage{amssymb}
\usepackage{multirow}
\usepackage{multicol}
\usepackage{booktabs}
\usepackage{lastpage}
\usepackage{hyperref} 
\usepackage{graphicx}
\usepackage{enumerate}
\usepackage{threeparttable}
\usepackage{float}
\usepackage{array}
\usepackage{cite}
\usepackage{color,soul}
\makeatletter
\renewcommand\section{\@startsection{section}{1}{\z@}%
  {-2.5ex \@plus -1ex \@minus -.2ex}%
  {2.3ex \@plus.2ex}%
  {\normalfont\large\bfseries}}
\renewcommand\subsection{\@startsection{subsection}{1}{\z@}%
  {-2.5ex \@plus -1ex \@minus -.2ex}%
  {2.3ex \@plus.2ex}%
  {\small\bfseries}}
\begin{document}
\title{\vskip -2.5 cm\textbf{Particle generation through restrictive planes in GEANT4 simulations for potential applications of cosmic ray muon tomography}}
\medskip
\author{\small A. Ilker Topuz$^{1,2}$, Madis Kiisk$^{1,3}$, Andrea Giammanco$^{2}$}
\medskip
\date{\small$^1$Institute of Physics, University of Tartu, W. Ostwaldi 1, 50411, Tartu, Estonia\\
$^2$Centre for Cosmology, Particle Physics and Phenomenology, Universit\'e catholique de Louvain, Chemin du Cyclotron 2, B-1348 Louvain-la-Neuve, Belgium\\
$^3$GScan OU, Maealuse 2/1, 12618 Tallinn, Estonia}
\maketitle
\begin{abstract}
In this study, by attempting to resolve the angular complication during the particle generation for the muon tomography applications in the GEANT4 simulations, we exhibit an unconventional methodology that is hinged on the direction limitation via the vectorial construction from the generation location to the restriction area rather than using a certain angular distribution or interval. In other words, we favor a momentum direction that is determined by a vector constructed between an initial point randomly chosen on a generative point/plane and a latter point arbitrarily selected on a restrictive plane of the same dimensions with the basal cross section of the volume-of-interest (VOI). On account of setting out such a generation scheme, we optimize the particle loss by keeping an angular disparity that is directly dependent on the VOI geometry as well as the vertical position of the restrictive plane for a tomographic system of a finite size. We demonstrate our strategy for a set of target materials including aluminum, copper, iron, lead, and uranium with a dimension of 40$\times$10$\times$40 $\rm cm^{3}$ over three restrictive planes of different positions by using a discrete energy spectrum between 0.1 and 8 GeV and we compute the scattering angle, the number of absorption, and the particle loss. Upon our simulation outcomes, we show that the particle generation by means of restrictive planes is an effective strategy that is flexible towards a variety of computational objectives in the GEANT4 simulations.
\end{abstract}
\section{Introduction}
The wide angular distribution~\cite{yanez2021method} of the incoming cosmic ray muons in connection with either incident angle or azimuthal angle is a challenging trait that leads to a drastic particle loss in the course of parametric computations through the GEANT4~\cite{agostinelli2003geant4} simulations associated with the muon tomography~\cite{pesente2009first, procureur2018muon, bonechi2020atmospheric} since the tomographic configurations as well as the target geometries also influence the processable number of the detected particles apart from the generation strategies. To further detail, the basic parameters such as the scattering angle, the particle displacement, and the particle absorption owing to the volume-of-interest (VOI) {\it{de facto}} dictate the particle penetration through the multiple sections of the tomographic setup in addition to the VOI. Hence, a number of the loss cases notably come into effect unless the calculation conditions are fulfilled, and not only the computation statistics as well as the numerical outcomes but the initial assumptions like the energy spectrum are also perturbed since the VOI accepts a significantly lower number of particles in the instance of the substantial particle loss. While a number of source biasing techniques~\cite{shultis2011mcnp} are offered by MCNP6~\cite{forster1985mcnp, goorley2012initial} in the black box format under the class of non-analogue Monte Carlo simulations, the GEANT4 simulations are usually constrained to the existing particle generators or the general particle source (GPS) unless G4ParticleGun is favored. Motivated by the excessive particle loss and its effect on the computation time as well as the characteristic parameters identified in the muon tomography, we set forth in the present study a scheme that is hinged on the particle generation through the planar restriction by means of the vectorial construction over our tomographic setup consisting of plastic scintillators manufactured from polyvinyl toluene with the dimensions of 100$\times$0.4$\times$100 $\rm cm^{3}$. This study is organized as follows. In section~\ref{methodology}, we elucidate our methodology based on the restrictive planes and we express our characteristic parameters as well as our simulation features in section~\ref{characteristicandsetup}. While we disclose our simulation outcomes in section~\ref{outcomes}, we draw our conclusions in section~\ref{conclusion}.
\section{Generation via planar restriction}
\label{methodology}
\begin{figure}[H]
\begin{center}
\includegraphics[width=7.7cm]{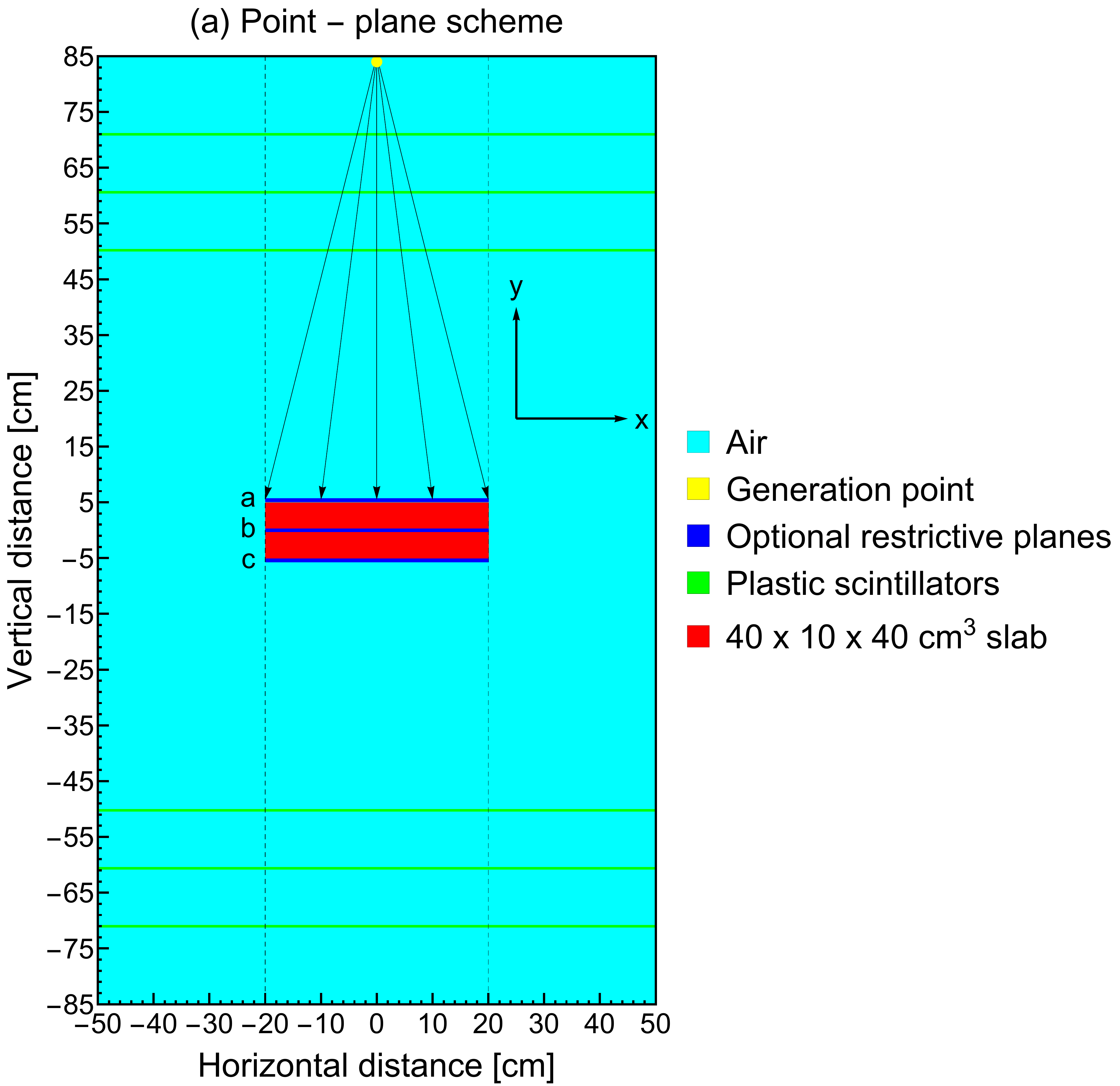}
\hskip 0.25cm
\includegraphics[width=7.7cm]{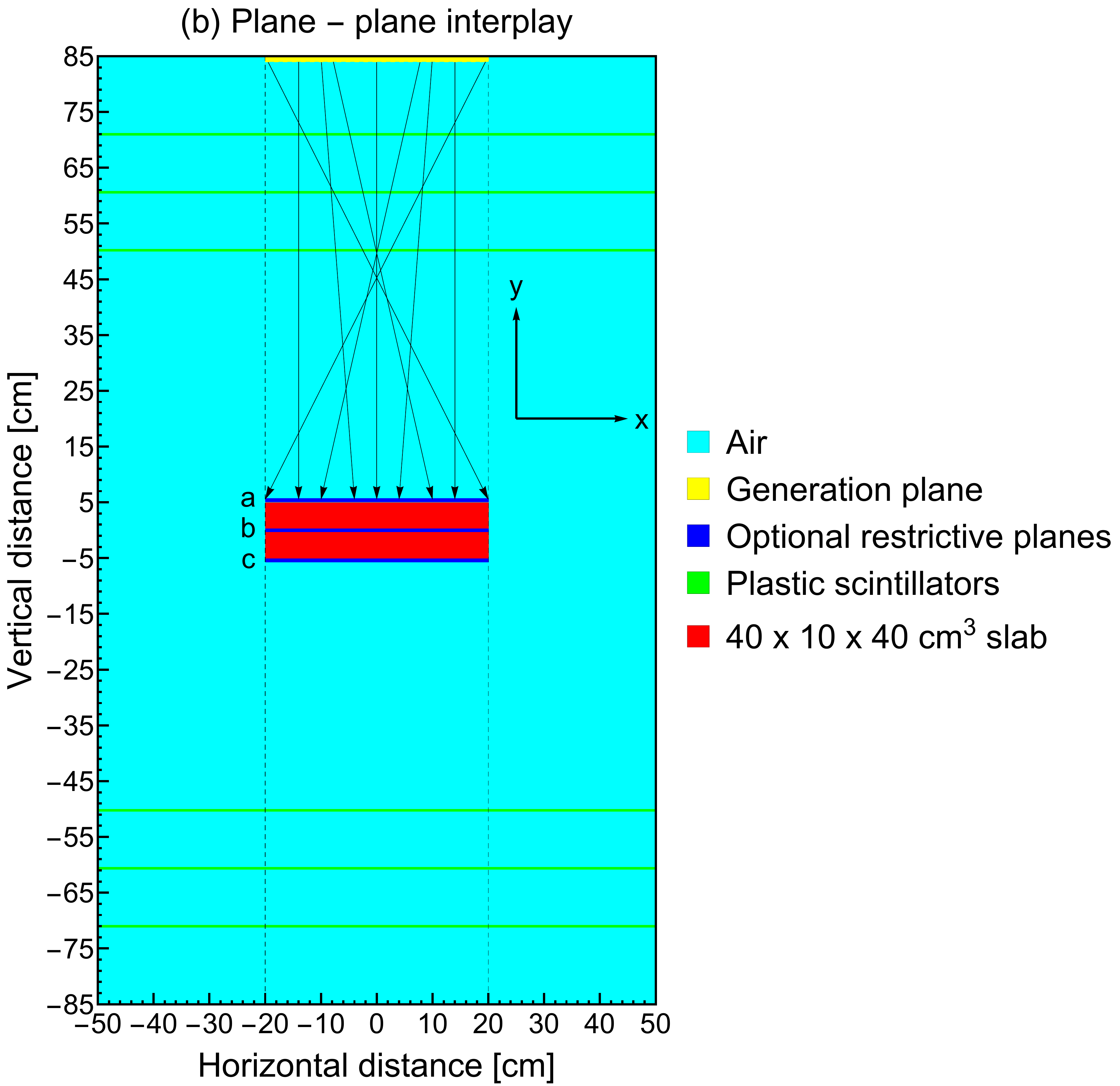}
\caption{Depiction of particle generation through restrictive planes in GEANT4 (a) generative point - restrictive plane scheme and (b) generative - restrictive planar interplay.}
\label{schemes}
\end{center}
\end{figure}

\begin{figure}[H]
\begin{center}
\includegraphics[width=6.45cm]{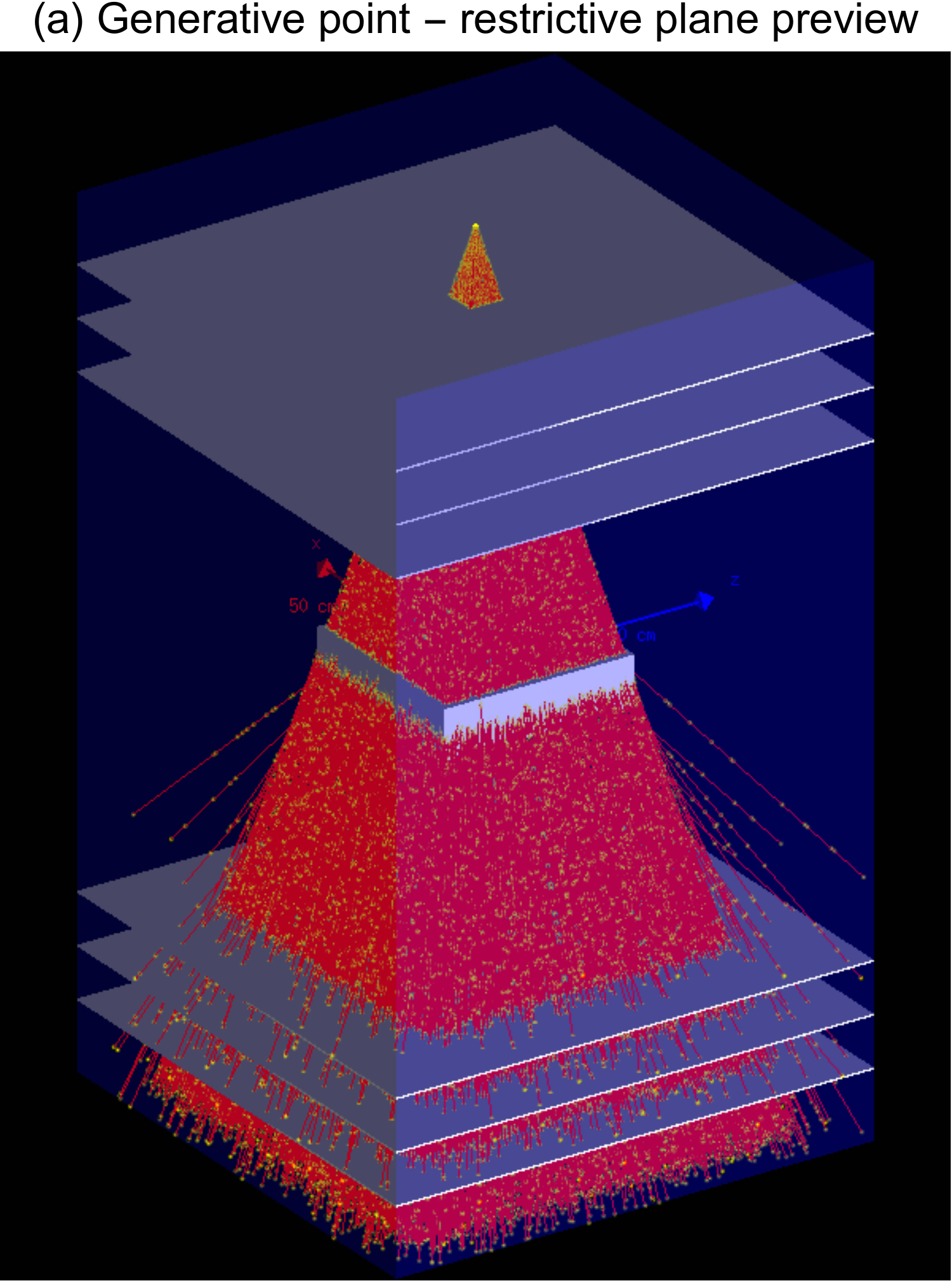}
\hskip 1cm
\includegraphics[width=6.5cm]{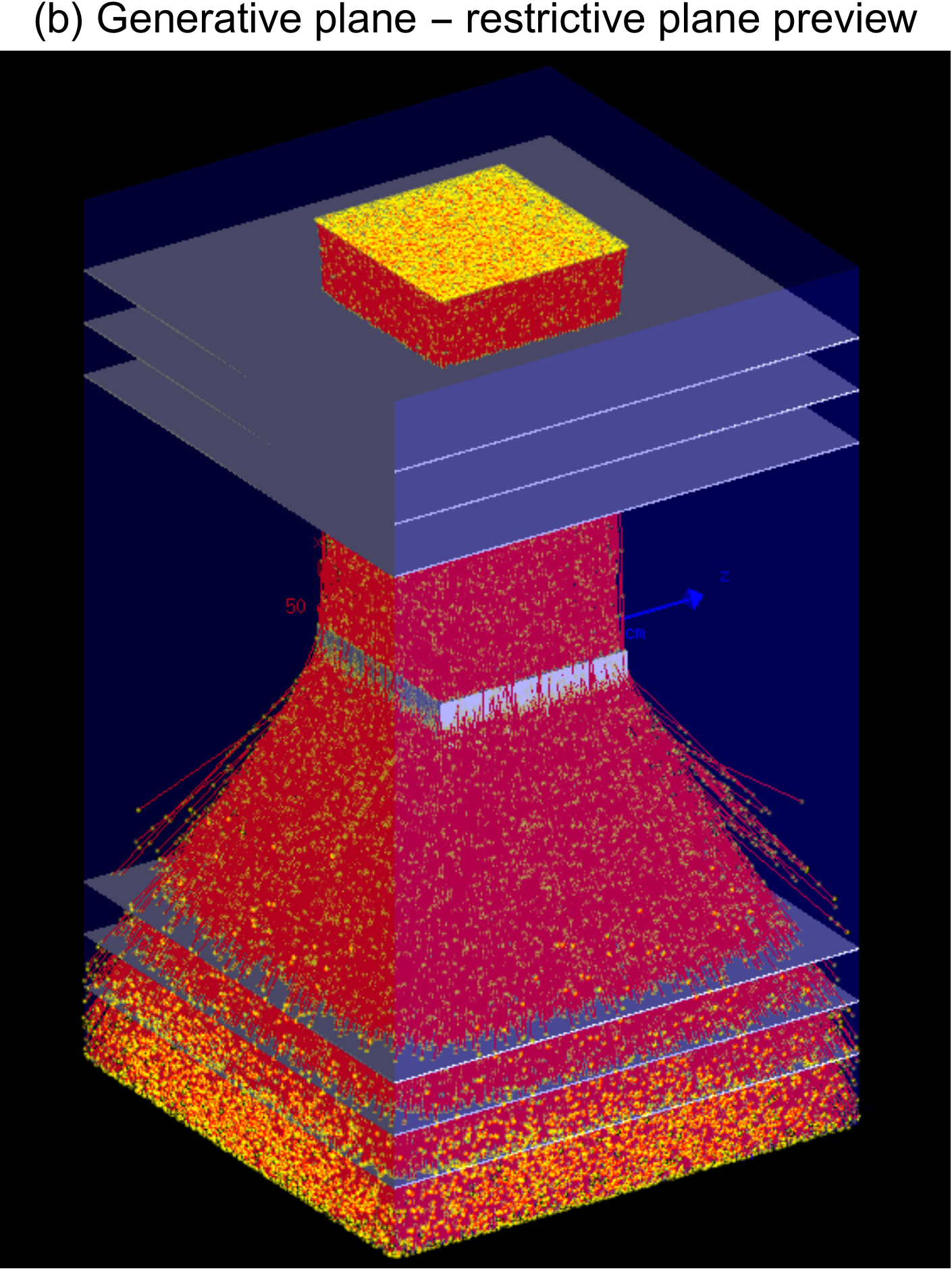}
\caption{Simulation previews by using restrictive plane b for copper in GEANT4 (a) point - plane scheme and (b) plane - plane scheme.}
\label{generation}
\end{center}
\end{figure}

To begin with, we principally exhibit two planar restrictive schemes to be adapted in GEANT4 as illustrated in Fig.~\ref{schemes} where (a) shows the particle generation from a fixed point as well as the direction restriction by means of a restrictive pseudo-plane, whereas (b) demonstrates the randomly picked up particles from a generative plane, the directions of which are projected into a similar restrictive plane.  
\vskip -0.75cm
In order to practically outline the present methodology that is initially described in Fig.~\ref{schemes}(a), the particle location in cm on the central point at height=85 cm is listed as written in
\begin{equation}
x_{0}=0,~~~
y_{0}=85,~~~
z_{0}=0
\end{equation}
Subsequently, the confined location in cm on any restrictive plane of $2L\times 2D$ $\rm cm^{2}$ is noted as shown in 
\begin{equation}
x_{1}=-L+2\times L\times{\rm G4UniformRand()},~~~
y_{1}={\rm constant},~~~
z_{1}=-D+2\times D\times{\rm G4UniformRand()}
\label{locationsonplane}
\end{equation}
Here, G4UniformRand() is the uniform random number generator between 0 and 1, which is pre-defined in GEANT4. Then, by constructing a vector from the generative point to the restrictive plane, we obtain
\begin{equation}
px=x_{1}-x_{0}=x_{1},~~~
py=y_{1}-y_{0},~~~
pz=z_{1}-z_{0}=z_{1}
\end{equation}
Thus, the selective momentum direction, i.e. $\vec{P}=(P_{x}, P_{y}, P_{z})$, is
\begin{equation}
P_{x}=\frac{px}{\sqrt{px^{2}+py^{2}+pz^{2}}},~~~
P_{y}=\frac{py}{\sqrt{px^{2}+py^{2}+pz^{2}}},~~~
P_{z}=\frac{pz}{\sqrt{px^{2}+py^{2}+pz^{2}}}
\end{equation}
The latter scheme that assumes a planar generation as delineated in Fig.~\ref{schemes}(b) entails particle locations in cm on the generative plane of $2L\times 2D$ $\rm cm^{2}$ as written in
\begin{equation}
x_{0}=-L+2\times L\times{\rm G4UniformRand()},~~~
y_{0}=85,~~~
z_{0}=-D+2\times D\times{\rm G4UniformRand()}
\end{equation}
As performed in Eq.~\ref{locationsonplane} for the previous scheme, the limited locations in cm on any restrictive plane of $2L\times 2D$ $\rm cm^{2}$ are selected from
\begin{equation}
x_{1}=-L+2\times L\times{\rm G4UniformRand()},~~~
y_{1}={\rm constant},~~~
z_{1}=-D+2\times D\times{\rm G4UniformRand()}
\end{equation}
Additionally, via a vector construction between two planes, we acquire anew
\begin{equation}
px=x_{1}-x_{0},~~~
py=y_{1}-y_{0},~~~
pz=z_{1}-z_{0}
\end{equation}
Therefore, the selective momentum direction denoted by $\vec{P}=(P_{x}, P_{y}, P_{z})$ is again
\begin{equation}
P_{x}=\frac{px}{\sqrt{px^{2}+py^{2}+pz^{2}}},~~~
P_{y}=\frac{py}{\sqrt{px^{2}+py^{2}+pz^{2}}},~~~
P_{z}=\frac{pz}{\sqrt{px^{2}+py^{2}+pz^{2}}}
\end{equation}
The initial particle positions and the selective momentum directions are incorporated by using G4ParticleGun. The simulation previews through both the restrictive schemes are displayed in Fig.~\ref{generation} where (a) indicates the particles generated from a fixed point, while (b) presents the randomly generated particles from a fixed plane.
 It is worth mentioning that neither generation points/planes nor restrictive planes are subject to any limitation in terms of shape, size, or location since our recent concept is preferred in the first instance for the sake of simplicity. On top of this, it is also possible to favor different distributions especially already implemented in GEANT4, e.g. Gauss or Poisson distribution depending on the envisaged application. 
\section{Characteristic parameters and simulation setup}
\label{characteristicandsetup}
Before getting down to test our schemes, we express our characteristic parameters to be computed in the wake of the GEANT4 simulations. The average scattering angle due to the target volume and its standard deviation over $N$ number of the non-absorbed/non-decayed muons is determined as expressed in~\cite{carlisle2012multiple,nugent2017multiple,poulson2019application}
\begin{equation}
\bar{\theta}\pm\delta\theta=\frac{1}{N}\sum_{i=1}^{N}\theta_{i}\pm\sqrt{\frac{1}{N}\sum_{j=1}^{N}(\theta_{j}-\bar{\theta})^{2}}
\end{equation}
Additionally, the root-mean-square (RMS) of the scattering angle over $N$ number of the non-absorbed/non-decayed muons is calculated by using the following expression:
\begin{equation}
\theta_{\rm RMS}=\sqrt{\frac{1}{N}\sum_{i=1}^{N}\theta_{i}^{2}}
\end{equation}
Along with the scattering angle, we squarely track the number of the absorbed muons within the VOI as denoted in
\begin{equation}
\#_{\rm Capture}^{\rm In-target}=\mbox{\# of muMinusCaptureAtRest in VOI}
\end{equation}
Last but not least, we define the particle loss entitled off-target loss as follows
\begin{equation}
\#^{\rm Off-target}_{\rm Loss}\approx\underbrace{\#_{\rm Out-scattering}}_{\rm Characteristic}+\underbrace{\#_{\rm Decay}}_{\rm Negligible}+\underbrace{\#^{\rm Off-target}_{\rm Capture}}_{\rm Negligible}+\underbrace{\#^{\rm Initial}_{\rm Deflection}}_{\rm Occasional}
\end{equation}
where $\#_{\rm Out-scattering}$ is the number of the scattered muons from the VOI by leaking out of the tomographic device, $\#_{\rm Decay}$ is the negligible number of the decayed muons into electrons/positrons, $\#^{\rm Off-target}_{\rm Capture}$ is the insignificant number of the absorbed muons outside the VOI, and $\#^{\rm Initial}_{\rm Deflection}$ is the number of muons that miss the VOI only in the case of the wide beams, which occasionally occurs due to the barriers before the VOI despite the initial restricted orientation to the VOI boundary, i.e. the tiny deflection owing to the detector layers. 
\vskip -0.5cm
\begin{table}[H]
\begin{center}
\begin{footnotesize}
\caption{Simulation features.}
\begin{tabular}{*2c}
\toprule
\toprule
Particle & $\mbox{\textmu}^{-}$\\
Momentum direction & Restrictive downward\\
Beam geometry & Prismatic\\
Initial position (cm) & y=85\\
Particle injector & G4ParticleGun\\
Number of particles & $10^{5}$\\
Energy distribution & Non-linear discrete\\
Energy interval & [0, 8]\\
Enegy bin step length (GeV) & 0.1\\
Target geometry & Rectangular prism\\
Target volume (cm$^{3}$) & 40$\times$10$\times$40\\ 
Material database & G4/NIST\\
Reference physics list & FTFP$\_$BERT\\
\bottomrule
\bottomrule
\label{features}
\end{tabular}
\end{footnotesize}
\end{center}
\end{table}
\vskip -1cm
Our simulation features are summarized in Table~\ref{features}, and we use a 80-bin discrete muon energy spectrum extracted from the CRY generator~\cite{hagmann2007cosmic} between 0 and 8 GeV. The muon tracking is accomplished by G4Step, and the recorded hit positions on the detector layers are post-processed at the hand of a Python script.
\section{Simulation outcomes}
\label{outcomes}
We asses our methodology over our tomographic configuration described in Fig.~\ref{schemes}(a)-(b) and we select our set of materials and the VOI geometry in accordance with another study~\cite{hohlmann2009geant4} dedicated to the muon tomography where the material list consists of aluminum, copper, iron, lead, and uranium, and the target geometry is composed of a rectangular prism with the dimensions of 40$\times$10$\times$40 $\rm cm^{3}$. As indicated in Fig.~\ref{schemes}, we contrast three restrictive planes labeled as a, b, and c that are placed atop the VOI, amidst the VOI, and beneath the VOI, respectively. We commence with the first scheme that is based on the point - plane generation, and the simulation outcomes by using restrictive plane a are listed in Table~\ref{pointplanea}. 
\vskip -0.6cm
\begin{table}[H]
\begin{center}
\begin{footnotesize}
\caption{Point - plane scheme, restrictive plane a, thickness=10 cm.}
\resizebox{0.75\textwidth}{!}{\begin{tabular}{*5c}
\toprule
\toprule
Material & $\bar{\theta} \pm\delta\theta$ [mrad] & $\theta_{\rm RMS} $ [mrad]& $\#^{\rm In-target}_{\rm Capture}$ & $\#^{\rm Off-target}_{\rm Loss}$\\
\midrule
Aluminum&14.890$\pm$25.741&29.738&-&516\\
Copper&37.376$\pm$55.515&66.924&1083&616\\
Iron&32.980$\pm$47.420&57.761&1073&541\\
Lead&59.486$\pm$81.898&101.222&1135&1215\\
Uranium&73.649$\pm$91.114&117.158&3267&1542\\
\bottomrule
\bottomrule
\label{pointplanea}
\end{tabular}}
\end{footnotesize}
\end{center}
\end{table}
\vskip -0.8cm
As shown in Table~\ref{pointplanea}, the computed parameters including the particle loss show a characteristic tendency depending on the atomic number as well as the material density for a fixed thickness. Although the muon beam is already directed to the VOI boundary even in the case of restrictive plane a, which leads to an immoderate reduction in the particle loss compared to the conventional approaches, a remarkable number of the loss events in agreement with the intrinsic properties of the target material are still observed.
\vskip -0.6cm  
\begin{table}[H]
\begin{center}
\begin{footnotesize}
\caption{Point - plane scheme, restrictive plane b, thickness=10 cm.}
\resizebox{0.75\textwidth}{!}{\begin{tabular}{*5c}
\toprule
\toprule
Material & $\bar{\theta} \pm\delta\theta$ [mrad] & $\theta_{\rm RMS} $ [mrad]& $\#^{\rm In-target}_{\rm Capture}$ & $\#^{\rm Off-target}_{\rm Loss}$\\
\midrule
Aluminum&15.771$\pm$26.427&30.775&-&54\\
Copper&39.545$\pm$56.941&69.326&1179&216\\
Iron&35.306$\pm$50.117&61.304&1172&133\\
Lead&63.172$\pm$84.172&105.241&1220&833\\
Uranium&78.160$\pm$93.551&121.904&3604&1187\\
\bottomrule
\bottomrule
\label{pointplaneb}
\end{tabular}}
\end{footnotesize}
\end{center}
\end{table}
\vskip -0.8cm
In order to see the positional effect of the planar restriction, the simulation outcomes from restrictive plane b are tabulated in Table~\ref{pointplaneb}. In comparison with Table~\ref{pointplanea}, we observe that the characteristic parameters except the particle loss slightly change when the muon beam is narrowed by using restrictive plane b; however, the particle loss manifests a minimum reduction of $31\%$ as opposed to restrictive plane a. Whereas restrictive plane b is capable of diminishing the particle loss by a factor of order in certain cases, we still notice that the particle loss remains distinctive among the simulated materials.

By using restrictive plane c, we further decrease the incident angle and we obtain the simulation results as written down in Table~\ref{pointplanec}. In comparison with Table~\ref{pointplaneb}, restrictive plane c yields a minuscule change in terms of the characteristic parameters containing the particle loss, which also means that the variation rate of the characteristic parameters is expected to be insignificant beyond restrictive plane c.
\vskip -0.6cm   
\begin{table}[H]
\begin{center}
\begin{footnotesize}
\caption{Point - plane scheme, restrictive plane c, thickness=10 cm.}
\resizebox{0.75\textwidth}{!}{\begin{tabular}{*5c}
\toprule
\toprule
Material & $\bar{\theta} \pm\delta\theta$ [mrad] & $\theta_{\rm RMS} $ [mrad]& $\#^{\rm In-target}_{\rm Capture}$ & $\#^{\rm Off-target}_{\rm Loss}$\\
\midrule
Aluminum&16.142$\pm$27.368&31.774&-&35\\
Copper&40.355$\pm$58.022&70.676&1216&193\\
Iron&35.916$\pm$50.635&62.080&1215&107\\
Lead&64.542$\pm$85.965&107.497&1287&793\\
Uranium&79.700$\pm$96.102&124.850&3764&1059\\
\bottomrule
\bottomrule
\label{pointplanec}
\end{tabular}}
\end{footnotesize}
\end{center}
\end{table}
\vskip -0.8cm
It is noteworthy to mention that a partial transition from the particle loss to the particle absorption is perceptible according to Tables~\ref{pointplanea}-\ref{pointplanec} especially if the VOI material is a potent absorber since the low-energy muons that lead to the particle loss in the wide beams typically have the absorption potential when interacting with the VOI material in the narrow beams, which also means that a certain portion o the particle loss is converted into the particle absorption in the VOI material towards restrictive plane c.
\vskip -0.6cm   
\begin{table}[H]
\begin{center}
\begin{footnotesize}
\caption{Plane - plane scheme, restrictive plane a, thickness=10 cm.}
\resizebox{0.75\textwidth}{!}{\begin{tabular}{*5c}
\toprule
\toprule
Material & $\bar{\theta} \pm\delta\theta$ [mrad] & $\theta_{\rm RMS} $ [mrad]& $\#^{\rm In-target}_{\rm Capture}$ & $\#^{\rm Off-target}_{\rm Loss}$\\
\midrule
Aluminum&15.196$\pm$26.036&30.146&-&1196\\
Copper&37.454$\pm$55.612&67.049&1118&1728\\
Iron&33.375$\pm$48.047&58.502&1092&1575\\
Lead&59.927$\pm$83.320&102.633&1206&2624\\
Uranium&74.073$\pm$92.787&118.728&3352&3299 \\
\bottomrule
\bottomrule
\label{planeplanea}
\end{tabular}}
\end{footnotesize}
\end{center}
\end{table}
\vskip -0.8cm
In the next step, we continue with the plane - plane scheme, and Table~\ref{planeplanea} lists the simulation outcomes for restrictive plane a. In spite of the schematic change, we see that the characteristic parameters excluding the particle loss do not exhibit a significant difference. On the other hand, the particle loss via restrictive plane a within the plane - plane interplay results in the elevated values as displayed in Table~\ref{planeplanea} in contrast to Tables~\ref{pointplanea}-\ref{pointplanec}.
\vskip -0.6cm    
\begin{table}[H]
\begin{center}
\begin{footnotesize}
\caption{Plane - plane scheme, restrictive plane b, thickness=10 cm.}
\resizebox{0.75\textwidth}{!}{\begin{tabular}{*5c}
\toprule
\toprule
Material & $\bar{\theta}\pm\delta\theta$ [mrad] & $\theta_{\rm RMS} $ [mrad]& $\#^{\rm In-target}_{\rm Capture}$ & $\#^{\rm Off-target}_{\rm Loss}$\\
\midrule
Aluminum&16.103$\pm$27.566&31.925&-&138 \\
Copper&39.897$\pm$57.927&70.337&1220&581\\
Iron&35.380$\pm$50.142&61.367&1206&430\\
Lead&63.335$\pm$84.573&105.659&1327&1423\\
Uranium&78.399$\pm$94.631&122.888&3699&1926\\
\bottomrule
\bottomrule
\label{planeplaneb}
\end{tabular}}
\end{footnotesize}
\end{center}
\end{table}
\vskip -0.8cm
So as to demonstrate the impact of the spatial change in the planar restriction for this scheme, the simulation results via restrictive plane b are tabulated in Table~\ref{planeplaneb}, and we experience a similar trend compared to the point-plane scheme that induces a drastic diminution in the particle loss along with the tiny variations in the rest of the characteristic parameters. As a means to complete our quantitative investigation for the plane - plane scheme, the simulation results for restrictive plane c are listed in Table~\ref{planeplanec}, and we face a close trend as opposed to Table~\ref{pointplanec}, which also means that the reduction rate in the particle loss is moderated together with the very minor variations in the remaining characteristic parameters.
\vskip -0.6cm    
\begin{table}[H]
\begin{center}
\begin{footnotesize}
\caption{Plane - plane scheme, restrictive plane c, thickness=10 cm.}
\resizebox{0.75\textwidth}{!}{\begin{tabular}{*5c}
\toprule
\toprule
Material & $\bar{\theta}\pm\delta\theta$ [mrad] & $\theta_{\rm RMS}$ [mrad] & $\#^{\rm In-target}_{\rm Capture}$ & $\#^{\rm Off-target}_{\rm Loss}$\\
\midrule
Aluminum&16.279$\pm$27.365&31.841&-&88\\
Copper&40.386$\pm$57.627&70.370&1258&389\\
Iron&36.135$\pm$50.751&62.300&1249&263\\
Lead&64.517$\pm$86.095&107.586&1358&1164\\
Uranium&80.087$\pm$96.225&125.193&3833&1537\\
\bottomrule
\bottomrule
\label{planeplanec}
\end{tabular}}
\end{footnotesize}
\end{center}
\end{table}
\vskip -0.8cm
In the long run, our last simulations are devoted to investigate the thickness effect by solely using restrictive plane b since we aim at optimizing the particle loss with an ideal angular acceptance. Thus, Table~\ref{pointplaneb40} shows the characteristic parameters that are acquired by means of the point - plane scheme as well as restrictive plane b for a thickness of 40 cm with the same material group.
\vskip -0.6cm    
\begin{table}[H]
\begin{center}
\begin{footnotesize}
\caption{Point - plane scheme, restrictive plane b, thickness=40 cm.}
\resizebox{0.75\textwidth}{!}{\begin{tabular}{*5c}
\toprule
\toprule
Material & $\bar{\theta} \pm\delta\theta$ [mrad] & $\theta_{\rm RMS} $ [mrad]& $\#^{\rm In-target}_{\rm Capture}$ & $\#^{\rm Off-target}_{\rm Loss}$\\
\midrule
Aluminum&27.849$\pm$37.186&46.458&3046&93\\
Copper&65.133$\pm$75.969&100.068&11072&588\\
Iron&58.208$\pm$67.893&89.429&10365&528\\
Lead&102.566$\pm$112.951&152.570&11036&2210\\
Uranium&121.060$\pm$121.502&171.517&20371&3084\\
\bottomrule
\bottomrule
\label{pointplaneb40}
\end{tabular}}
\end{footnotesize}
\end{center}
\end{table}
\vskip -0.8cm
From Table~\ref{pointplaneb40}, we numerically demonstrate that all the characteristic parameters increase as a function of thickness, and we find the most notable rise in the particle absorption. Finally, Table~\ref{planeplaneb40} lists the simulation results through the plane-plane scheme for the same thickness, and we see that the latter scheme is not significantly different from the initial scheme with regard to the characteristic parameters omitting a higher number of the particle loss. 
\vskip -0.6cm    
\begin{table}[H]
\begin{center}
\begin{footnotesize}
\caption{Plane - plane scheme, restrictive plane b, thickness=40 cm.}
\resizebox{0.75\textwidth}{!}{\begin{tabular}{*5c}
\toprule
\toprule
Material & $\bar{\theta} \pm\delta\theta$ [mrad] & $\theta_{\rm RMS} $ [mrad]& $\#^{\rm In-target}_{\rm Capture}$ & $\#^{\rm Off-target}_{\rm Loss}$\\
\midrule
Aluminum&28.022$\pm$37.620&46.910&3080&272\\
Copper&65.229$\pm$77.147&101.026&11341&1184\\
Iron&58.373$\pm$68.363&89.894&10599&1086\\
Lead&101.906$\pm$113.230&152.335&11341&3341\\
Uranium&120.089$\pm$121.872&171.097&20867&4181\\
\bottomrule
\bottomrule
\label{planeplaneb40}
\end{tabular}}
\end{footnotesize}
\end{center}
\vspace{-1cm}
\end{table}
\section{Conclusion}
\label{conclusion}
All in all, by setting out our restrictive generation scheme, we optimize the particle loss by keeping an angular disparity that is directly dependent on the VOI geometry as well as the vertical position of the restrictive plane for a tomographic system of a finite size. Upon our simulation outcomes, we show that the particle generation by means of restrictive planes is an effective strategy that is flexible towards a variety of computational objectives in GEANT4. Into the bargain, we explicitly observe that the off-target loss is a characteristic parameter that varies in an ascending order from aluminum to uranium. 
\bibliographystyle{elsarticle-num}
\bibliography{Restrictive.bib}

\begin{thebibliography}{10}
\expandafter\ifx\csname url\endcsname\relax
  \def\url#1{\texttt{#1}}\fi
\expandafter\ifx\csname urlprefix\endcsname\relax\def\urlprefix{URL }\fi
\expandafter\ifx\csname href\endcsname\relax
  \def\href#1#2{#2} \def\path#1{#1}\fi

\bibitem{yanez2021method}
B.~O. Y{\'a}{\~n}ez, A.~A. Aguilar-Arevalo, A method to measure the integral
  vertical intensity and angular distribution of atmospheric muons with a
  stationary plastic scintillator bar detector, Nucl. Instr. Meth. A 987 (2021)
  164870.

\bibitem{agostinelli2003geant4}
S.~Agostinelli, J.~Allison, K.~Amako, J.~Apostolakis, H.~Araujo, P.~Arce,
  M.~Asai, D.~Axen, S.~Banerjee, G.~Barrand, et~al., {GEANT4—a simulation
  toolkit}, Nuclear instruments and methods in physics research section A:
  Accelerators, Spectrometers, Detectors and Associated Equipment 506~(3)
  (2003) 250--303.

\bibitem{pesente2009first}
S.~Pesente, S.~Vanini, M.~Benettoni, G.~Bonomi, P.~Calvini, P.~Checchia,
  E.~Conti, F.~Gonella, G.~Nebbia, S.~Squarcia, et~al., {First results on
  material identification and imaging with a large-volume muon tomography
  prototype}, Nuclear Instruments and Methods in Physics Research Section A:
  Accelerators, Spectrometers, Detectors and Associated Equipment 604~(3)
  (2009) 738--746.

\bibitem{procureur2018muon}
S.~Procureur, {Muon imaging: Principles, technologies and applications}, Nucl.
  Instr. Meth. A 878 (2018) 169.

\bibitem{bonechi2020atmospheric}
L.~Bonechi, R.~D’Alessandro, A.~Giammanco, Atmospheric muons as an imaging
  tool, Reviews in Physics 5 (2020) 100038.

\bibitem{shultis2011mcnp}
J.~K. Shultis, R.~E. Faw, {An MCNP primer}, Tech. rep., Manhattan, Kansas State
  University (2011).

\bibitem{forster1985mcnp}
R.~Forster, T.~Godfrey, {MCNP - a general Monte Carlo code for neutron and
  photon transport}, in: Monte Carlo Methods and Applications in Neutronics,
  Photonics and Statistical Physics, Springer, 1985, pp. 33--55.

\bibitem{goorley2012initial}
T.~Goorley, M.~James, T.~Booth, F.~Brown, J.~Bull, L.~Cox, J.~Durkee, J.~Elson,
  M.~Fensin, R.~Forster, et~al., {Initial MCNP6 release overview}, Nuclear
  technology 180~(3) (2012) 298--315.

\bibitem{carlisle2012multiple}
T.~Carlisle, J.~Cobb, D.~Neuffer, {Multiple Scattering Measurements in the MICE
  Experiment}, Tech. Rep. {FERMILAB-CONF-12-171-APC}, Fermi National
  Accelerator Lab. (FNAL), Batavia, IL (United States) (2012).

\bibitem{nugent2017multiple}
J.~C. Nugent, {Multiple Coulomb scattering in the MICE experiment}, Ph.D.
  thesis, University of Glasgow (2017).

\bibitem{poulson2019application}
D.~Poulson, J.~Bacon, M.~Durham, E.~Guardincerri, C.~Morris, H.~R. Trellue,
  Application of muon tomography to fuel cask monitoring, Philosophical
  Transactions of the Royal Society A 377~(2137) (2019) 20180052.

\bibitem{hagmann2007cosmic}
C.~Hagmann, D.~Lange, D.~Wright, {Cosmic-ray shower generator (CRY) for Monte
  Carlo transport codes}, in: 2007 IEEE Nuclear Science Symposium Conference
  Record, Vol.~2, IEEE, pp. 1143--1146.

\bibitem{hohlmann2009geant4}
M.~Hohlmann, P.~Ford, K.~Gnanvo, J.~Helsby, D.~Pena, R.~Hoch, D.~Mitra, {GEANT4
  Simulation of a Cosmic Ray Muon Tomography System With Micro-Pattern Gas
  Detectors for the Detection of High-Z Materials}, IEEE Transactions on
  Nuclear Science 56~(3) (2009) 1356--1363.

\end{thebibliography}
\end{document}